\documentclass[prx,showpacs,superscriptaddress,amssymb,twocolumn,notitlepage,nofootinbib,longbibliography]{revtex4-2}
\usepackage{graphicx}
\usepackage{amsmath}
\usepackage{amssymb}
\usepackage[usenames]{color}
\usepackage{hyperref}
\usepackage{placeins}

\hypersetup{
    colorlinks=true,       
    linkcolor=cyan,          
    citecolor=magenta,        
    filecolor=magentan           u8j,      
    urlcolor=cyan,           
    runcolor=cyan
}
\usepackage{subfigure}

\newcommand{\braket}[2]{\langle #1 | #2\rangle}
\newcommand{\ket}[1]{ | #1\rangle}
\newcommand{\bra}[1]{\langle #1 | }
\newcommand{\ketbra}[2]{|#1\rangle\langle #2|}
\newcommand{\beq}{\begin{equation}}
\newcommand{\eeq}{\end{equation}}
\newcommand{\beqnn}{\begin{equation*}}
\newcommand{\eeqnn}{\end{equation*}}
\newcommand{\bea}{\begin{eqnarray}}
\newcommand{\eea}{\end{eqnarray}}
\newcommand{\beann}{\begin{eqnarray*}}
\newcommand{\eeann}{\end{eqnarray*}}
\newcommand{\bes} {\begin{subequations}}
\newcommand{\ees} {\end{subequations}}

\newcommand{\ident}{\openone}
\usepackage{siunitx}
\usepackage{mathtools}

\begin{document}
\title{Optimizing Temperature Distributions for Training Neural Quantum States using Parallel Tempering}
\author{Conor Smith}
\affiliation{Center for Quantum Information and Control, University of New Mexico, University of New Mexico, Albuquerque, NM, USA}
\affiliation{Department of Electrical and Computer Engineering, University of New Mexico, Albuquerque, NM, USA}
\affiliation{Center for Computational Quantum Physics, Flatiron Institute, New York, NY, 10010, USA}
\author{Quinn T. Campbell}
\affiliation{Center for Computing Research, Sandia National Laboratories, Albuquerque NM, 87185 USA}
\author{Tameem Albash}
\affiliation{Center for Computing Research, Sandia National Laboratories, Albuquerque NM, 87185 USA}

\begin{abstract}
Parameterized artificial neural networks (ANNs) can be very expressive ansatzes for variational algorithms, reaching state-of-the-art energies on many quantum many-body Hamiltonians. Nevertheless, the training of the ANN can be slow and stymied by the presence of local minima in the parameter landscape. One approach to mitigate this issue is to use parallel tempering methods, and in this work we focus on the role played by the temperature distribution of the parallel tempering replicas. Using an adaptive method that adjusts the temperatures in order to equate the exchange probability between neighboring replicas, we show that this temperature optimization can significantly increase the success rate of the variational algorithm with negligible computational cost by eliminating bottlenecks in the replicas' random walk. We demonstrate this using two different neural networks, a restricted Boltzmann machine and a feedforward network, which we use to study a toy problem based on a permutation invariant Hamiltonian with a pernicious local minimum and the $J_1$-$J_2$ model on a rectangular lattice.

\end{abstract}
\maketitle
\section{Introduction}
%
Finding or approximating the ground state of a many-body quantum Hamiltonian is an optimization problem at the core of understanding many scientifically relevant systems, such as strongly-correlated materials, quantum chemistry, and high energy physics. A standard approach to tackling this problem is applying the variational method to a parameterized wavefunction. An exact representation is generally too costly at physically relevant system sizes, so a practical representation needs to be efficient in terms of the number of its parameters but expressive enough to capture the ground state to the desired accuracy. Furthermore, the representations should be paired with an efficient algorithm to perform updates to the parameterized wavefunction.  The approach is then practical if the solution is found in a total number of updates to the parameterized wavefunction that is not prohibitive. While we should not expect any single approach to be efficient for all Hamiltonian problems \cite{Kitaev2002,Kempe2003,Kempe2006,OGorman2021,Hua2021}, new representations and algorithmic developments can help grow the number and size of Hamiltonians that can be tackled using classical computing~\cite{Astrakhantsev2021,Wilson2021,Nomura2021,Lovato2022,Ren2023,Zhao2023,Zhou2023,Rende2024}.

The flexibility of artificial neural networks (ANNs) has made them an attractive choice for parameterized wavefunctions that can be trained using variational Monte Carlo (VMC) methods, such as stochastic reconfiguration (SR)~\cite{Sorella1998,Sorella2001} and minimum-step stochastic reconfiguration~\cite{Chen2023}, to approximate ground states~\cite{Carleo2017}. An ANN used to represent a quantum state is now commonly referred to as a Neural Quantum State (NQS). State-of-the-art results can be achieved by carefully choosing the ANN architecture to reproduce the symmetries of the Hamiltonian of interest\cite{Roth2021,Roth2023}. For recent reviews of NQSs and their applications, see Ref.~\cite{Jia2019,Lange2024,Medvidovic2024}. Despite the success and promise of NQSs, training these networks to approximate the ground state can still be resource intensive \cite{Liang2023}, with the primary culprits being frustration and the sign structure of the ground state~\cite{Westerhout2020,Szabo2020,Bukov2021}.

In this work, we continue the development of using parallel tempering (PT)~\cite{Swendsen1986,Geyer1991,Hukushima1996} as a metaheuristic for training NQSs~\cite{Albash2023}. By simulating multiple systems in parallel, with each operating at a different temperature, PT aims to overcome barriers in the energy landscape by allowing systems to be kicked to higher temperatures where they can escape local minima and then relax back to colder temperatures and reach lower energy configurations. Here we focus on the role played by the temperature distribution of the replicas, known to play an important role in the efficiency of PT in the study of spin glasses~\cite{Katzgraber2006}. By using an adaptive method for optimizing the temperature distribution of replicas, we demonstrate an improvement in the success frequency beyond what is achieved with PT without the adaptive temperature.

We emphasize that training with PT does not improve the expressivity of the NQS used, so the lowest energy that could in principle be achieved is not changed. Success of the approach is instead measured in terms of how long it takes the training to reach a desired level of accuracy. For this reason, our work focuses on the number of updates it takes to reach a target energy for the same NQS, and we do not use state-of-the-art NQS's that can achieve the lowest known energies. In practice, choosing the NQS that most efficiently approximates the ground state and the training algorithm that most effectively reaches that approximation should go hand in hand.

The manuscript is organized as follows. In Sec.~\ref{sec:setup}, we describe our methodology, including the ANN architectures, variational algorithms, and PT method used. In Sec.~\ref{sec:results}, we present our results for the two classes of problems we study, the Precipice problem and the $J_1$-$J_2$ model.  The Precipice problem has an energy landscape that drives the training to a local minimum from which it is very difficult to escape and serves as a useful toy model to demonstrate how our temperature optimization can improve performance. Because of its simplicitly, we study this model with only a Restricted Boltzmann Machine (RBM) \cite{Smolensky1986,Hinton2002}. For the $J_1$-$J_2$ model, we use both an RBM and a feedforward~\cite{Hornik1989} network architecture in order to illustrate the usefulness of our methods on different architectures. In Sec.~\ref{sec:conclusions}, we discuss our results and conclusions.
\section{The setup} \label{sec:setup}
%
\subsection{Wavefunction Representation}
We define our (unnormalized) parameterized wavefunction $\ket{\psi(\alpha)}$ as:
\beq
\ket{\psi(\alpha)} = \sum_{x \in \left\{0,1\right\}^{\otimes n} } \psi_x(\alpha) \ket{x} \ ,
\eeq
where the set $\left\{ \ket{x} \right\}_x$ is the computational basis for $n$ qubits\footnote{We pick the convention that for the Pauli $z$ operator on a single qubit, the computational basis satisfies $\sigma^z \ket{0} = \ket{0}$ and $\sigma^z \ket{1} = - \ket{1}$.}. The function $\psi_x(\alpha)$ is the output of an ANN with parameters $\left\{ \alpha_k \right\}_k$ given the input $x$. In this work, we focus on two types of ANN's, the Restricted Boltzmann Machine (RBM) and a three layer feedforward neural network. For an RBM with $n$ visible nodes and $m$ hidden nodes, the parameterized function is given by:
\beq
\psi_x(\alpha) = e^{\sum_{i=1}^n a_i z_i} \prod_{\mu=1}^m \cosh\left( b_\mu + \sum_{k=1^n} W_{\mu k} z_k \right) \ ,
\eeq
where $z_i = 1-2x_i$ and where we have expressed the $(n+m + nm)$ complex parameters $\alpha$ in terms of the complex set $\left\{a_i, b_\mu, W_{\mu k} \right\}$. 

On larger Hamiltonian systems with a more complex ground state structure, it is advantageous to use a deeper network, so we add two layers to the above construction to form our three layer feedforward ANN. In this case, the output of the ANN can be expressed as
\begin{equation}
    \psi_x(\alpha) = \phi^{(2)}_{h}(\phi^{(1)}_{h}(\phi_{v}(z_i))) \ ,
\end{equation}
where $\phi_{v(h)} = \sigma(W^{v(h)}_{ji}z_i)$ is the visible (hidden) layer with the learnable weight matrix $W^{v(h)}$ and activation function $\sigma$.  All activation functions are Gaussian Error Linear Unit (GELU) functions~\cite{Hendrycks2016} defined as:
\begin{equation}
    \sigma(x) = \frac{x}{2}\left(1+\mathrm{erf}\left(\frac{x}{\sqrt{2}}\right) \right) \ ,
\end{equation}
where $\mathrm{erf}(x)$ is the Gauss error function.

The RBM construction allows for outputs and gradients to be computed directly. For the feedforward and more general structures, auto-differentiation is convenient to compute gradients. We therefore implement the feedforward network in the NQS VMC package NetKet~\cite{netket2:2019,netket3:2022}, which is built on JAX~\cite{jax2018github} and allows the use of arbitrary network structures.
\subsection{Optimizing the Free Energy} \label{sec:SR}
%
In our work, the target of the training of the ANN's is to minimize a cost function that is akin to the free energy
\beq \label{eqt:FreeEnergy}
F(\alpha,\alpha^\ast) = \frac{\bra{ \psi(\alpha) } \hat{H} \ket{\psi(\alpha)}}{ \braket{\psi(\alpha)}{\psi(\alpha)}} - T  S(\alpha,\alpha^\ast) \ ,
\eeq
where $T$ is the temperature of the replica and $S(\alpha,\alpha^\ast)$ is the von-Neumann entropy of the probability distribution associated with measurements in the computational basis:
\beq
 S (\alpha,\alpha^\ast)  = - \sum_x p_x(\alpha,\alpha^\ast) \ln p_x(\alpha,\alpha^\ast) \ ,
\eeq
where $p_x(\alpha,\alpha^\ast) =  \frac{|\braket{x}{\psi(\alpha)}|^2}{\braket{\psi(\alpha)}{\psi(\alpha)}}$.
Training with this cost function was used in Refs.~\cite{Hibat2022,Roth2023} in the context of annealing but here we use it in the context of our PT scheme (see also Ref.~\cite{Albash2023} for an alternative to the free energy). 

To see how introducing the entropy term to the cost function changes the update rule for SR (see Appendix A of Ref.~\cite{Albash2023} for a derivation), we can expand Eq.~\eqref{eqt:FreeEnergy} to first order in the perturbations to the parameters $\alpha$ and $\alpha^\ast$. We have:
\begin{eqnarray}
F(\alpha + \delta \alpha, \alpha^\ast + \delta \alpha^\ast)
&=& F(\alpha,\alpha^\ast ) \nonumber \\
&& \hspace{-3.5cm} + \sum_k\left(  \delta \alpha_k^\ast  \frac{\bra{\psi(\alpha)} \left( \hat{O}_k^\dagger - \langle \hat{O}_k^\dagger \rangle \right) \hat{F}(\alpha,\alpha^\ast)  \ket{\psi(\alpha)}}{{ \braket{\psi(\alpha)}{\psi(\alpha)}}}  \right. \nonumber  \\
&& \hspace{-3.5cm} +  \delta \alpha_k  \frac{\bra{\psi(\alpha)} \hat{F}(\alpha,\alpha^\ast)\left( \hat{O}_k - \langle \hat{O}_k \rangle \right)  \ket{\psi(\alpha)}}{ \braket{\psi(\alpha)}{\psi(\alpha)}} + \dots \ ,
\end{eqnarray}
where we defined the parameter-dependent operators:
\bes
\begin{align}
\hat{F}(\alpha,\alpha^\ast)&= \hat{H} - T \hat{S}(\alpha,\alpha^\ast) \ , \\
\hat{S}(\alpha, \alpha^\ast) &= - \sum_x \ln ( |\psi_x(\alpha)|^2) \ketbra{x}{x} \ , \\
\hat{O}_k(\alpha) &= \sum_x  \frac{1}{\psi_x(\alpha)} \frac{\partial}{\partial \alpha_k} \psi_x(\alpha) \ketbra{x}{x} \ , 
\end{align}
\ees
and the expectation value of an operator $\hat{A}$ is defined as:
\beq
\langle \hat{A} \rangle = \frac{\bra{\psi(\alpha)} \hat{A} \ket{\psi(\alpha)}}{\braket{\psi(\alpha)}{\psi(\alpha)}} \ .
\eeq
Minimizing the free energy $F(\alpha,\alpha^\ast) $ using SR amounts to performing the updates
\beq \label{eqt:SRupdate}
\delta \alpha_k = - \eta \sum_{k'} s_{kk'}^{-1} f_{k'} \ ,
\eeq
where $\eta$ is the learning rate and where the force vector $f_k$ and covariance matrix $s_{kk'}$ are defined as (suppressing the dependence on $\alpha, \alpha^\ast$):
\bes \label{eqt:sandf}
\begin{align}
f_k &= 
 \langle \hat{O}_k^\dagger \hat{F}  \rangle - \langle \hat{O}_k^\dagger \rangle \langle \hat{F} \rangle \ ,\\
s_{kk'} &=
 \langle \hat{O}_k^\dagger \hat{O}_{k'} \rangle - \langle \hat{O}_k^\dagger \rangle \langle \hat{O}_{k'} \rangle \ .
\end{align}
\ees
Thus the introduction of the entropy term amounts to replacing the local energy that appears in the force vector with the local energy plus $T \ln( |\psi_x(\alpha)|^2)$~\cite{Roth2023}.
%
\subsection{Parallel Tempering and Optimizing the swap probability} \label{sec:opt-swap-prob}
%
In order to help the training of our ANN's escape from local minima, we implement 
PT~\cite{Swendsen1986,Geyer1991,Hukushima1996} as part of our optimization. Details of the adaptation of the method to training ANN's are provided in Ref.~\cite{Albash2023}, but we provide a summary here for completeness.

Multiple ANN's, each called a replica, are trained in parallel with SR.  The $i$-th replica is trained using a unique temperature $T_i$ as described in Sec.~\ref{sec:SR}, with the 0-th replica set at zero temperature.  After a fix number of SR updates, a configuration swap of neighboring replicas is attempted with probability
\begin{eqnarray} \label{eqt:SwapProbability}
p_{i,i+1} &=& \min \left( 1, \exp \left[ \left( \beta_i - \beta_{i+1} \right) \left( \overline{E}_i - \overline{E}_{i+1} \right) \right] \right) ,
\end{eqnarray}
for $i \in [0, n_{\mathrm{R}}-2]$ and where $\beta_i = 1/T_i$ is the inverse temperature and $\overline{E}_i$ is a running average of the expectation value $\langle \hat{H} \rangle$.  In our implementation of PT, we attempt neighbor swaps between $i$ even and $i+1$ odd replicas after $n_{\mathrm{swap}}$ SR updates, and then attempt neighbor swaps between $i$ odd and $i+1$ even replicas after the next $n_{\mathrm{swap}}$ SR updates.  This particular choice is purely one of convenience since it allows us to implement the swap updates on the different pairs of replicas in parallel.

The aim of using PT is to allow configurations of the ANN that have been trained at high temperature to be transferred with some probability to lower temperature, which can help lower temperature replicas to be bumped out of local minima of the training landscape. The PT approach becomes ineffective when the swap probability (Eq.~\eqref{eqt:SwapProbability}) between neighboring replicas is very low, effectively preventing replicas on one side of the temperature distribution from mixing with replicas on the other side. This problem can be mitigated by introducing additional replicas between the problematic temperatures and by optimizing the temperature distribution to avoid such bottlenecks.

Our strategy to optimize the temperature distribution is to pick an inverse-temperature distribution to ensure that we have equal swap probabilities between replicas \cite{Hukushima1999,Rozada2019}.
We propose to replace $\beta_i$ for $i \in [1, n_{\mathrm{R}} - 2]$ (excludes the lowest and highest temperatures) by:
\beq \label{eqt:temperatureUpdate}
 \beta_i^\ast = \frac{1}{2} \left( \beta_i + \frac{\beta_{i-1} \overline{E}_- + \beta_{i+1} \overline{E}_+}{\overline{E}_- + \overline{E}_+}\right) \ ,
\eeq
where $\overline{E}_{-} = \overline{E}_{i-1} - \overline{E}_i, \overline{E}_+ = \overline{E}_i - \overline{E}_{i+1}$.  The update is accepted if $\beta_{i-1} > \beta_i^{\ast} > \beta_{i+1}$. The second term in the above expression arises from demanding that the probability of swapping between the $i-1 \leftrightarrow i$ replicas (as defined in Eq.~\ref{eqt:SwapProbability}) and the $i \leftrightarrow i+1$ replicas be equal, i.e.:
\beq
\left(\beta_{i-1} - \beta_i \right) \left( \overline{E}_{i-1} - \overline{E}_{i} \right) = \left(\beta_i - \beta_{i+1} \right) \left( \overline{E}_{i} - \overline{E}_{i+1} \right).
\eeq
To prevent rapid changes in the inverse-temperature $\beta_i^\ast$, we take an average of its current value $\beta_i$ and the solution to the above equation. Our approach is to first optimize the odd values of $i$ followed by the even values \cite{Rozada2019}.

\section{Results} \label{sec:results}
%
\subsection{Precipice problem} \label{sec:Precipice}
%
In order to illustrate the advantages afforded by optimizing the temperature distribution, we start by studying a toy model with a deep local minimum in its energy landscape. Our target Hamiltonian is given by the ``Precipice problem'' \cite{vanDam2001} with a uniform transverse field:
\beq \label{eqt:PrecipiceH}
\hat{H} = \frac{1}{2}(1-s) \sum_{i=1}^n \left( \hat{\ident} - \hat{\sigma}_i^x \right) + s \sum_{x \in \left\{ 0 , 1 \right\}^{\otimes n} } f(x) \ketbra{x}{x} \ ,
\eeq
where $f(x)$ is a function that only depends on the Hamming weight of the $n$-bit string $x$ and is given by:
\beq
f(x) = \left\{ \begin{array} {lr}
-1 & \text{if }|x| = n \\
|x| & \text{otherwise}
\end{array} \right. \ . 
\eeq
We pick $s = 4/5$ so that the Hamiltonian $\hat{H}$ is dominated by the contribution from $f(x)$, which exhibits a local minimum at $x=0$ that is a Hamming distance of $n$ from the global minimum at $x=2^{n} - 1$. The local minimum is pernicious because after initializing the wavefunction randomly the energy gradient calculated by SR pushes the ANN wavefunction towards the local minimum with high probability.  Only if the wavefunction happens to start with a sufficient overlap with the global minimum will the wavefunction be pushed towards it with SR. Once in the local minimum, it is extremely unlikely to escape unless an appropriate escape mechanism such as PT is used \cite{Albash2023}.

The precipice model serves as a a useful test case because we can easily eliminate the role of ANN expressibility and sampling from the effectiveness of the training and hence study the role played by PT. The Hamiltonian $\hat{H}$ is qubit permutation invariant, and its ground state is symmetric under qubit permutations. Therefore, we can restrict the RBM ansatz to satisfy these conditions by having it depend only on the Hamming weight of the input $x$:
\beq
\psi_x(\alpha) = e^{a \sum_{i=1}^n z_i} \prod_{\mu = 1}^m \cosh \left( b_\mu + W_\mu \sum_{i=1}^n z_i \right) \ , 
\eeq
which dramatically reduce the number of variational parameters to $1 + 2m$. This allows us to consider large choices of $m$ with little computational cost to ensure the RBM is expressive enough to describe the ground state of $\hat{H}$.  Furthermore, since all computational basis states with the same Hamming weight have the same amplitude, we can perform exact sampling \emph{without} an exponential cost in the system size $n$. Details of the simulation settings are provided in Appendix~\ref{app:PrecipiceSettings}.

We study the case of $n=32$ with a number of replicas $n_{\mathrm{R}}$ in the range $[6,10]$. In our previous study \cite{Albash2023}, this number of replicas was insufficient to find the ground state for this system size, but as we show here becomes sufficient for $n_{\mathrm{R}} \geq 7$ by optimizing the temperature distribution.

\begin{figure}
    \centering
  \subfigure[Without optimization]{\includegraphics[width=2.95in]{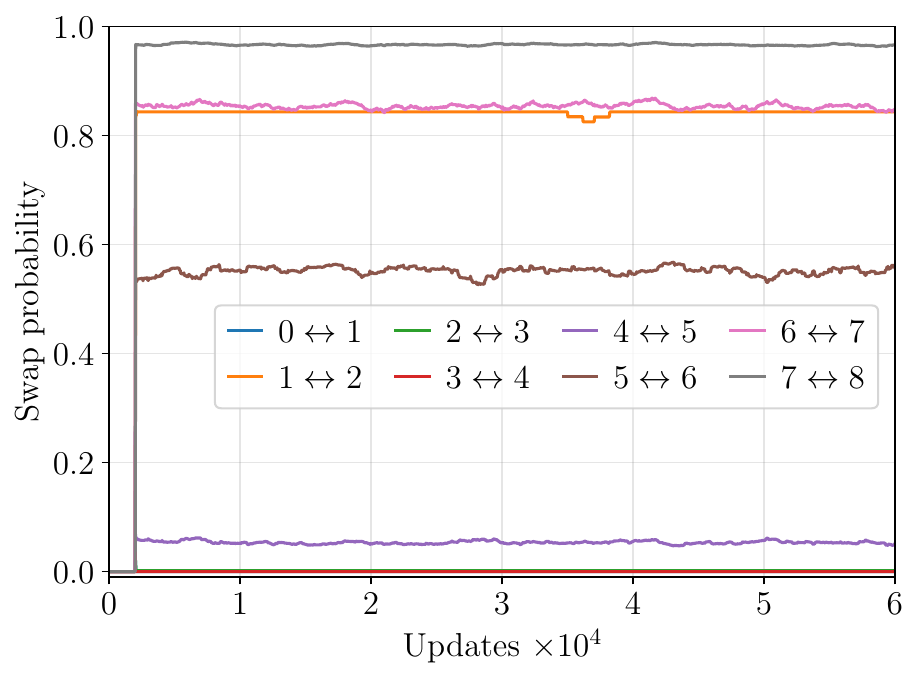} \label{fig:precipice-swap-probs-without}}
  \subfigure[With optimization]{\includegraphics[width=2.95in]{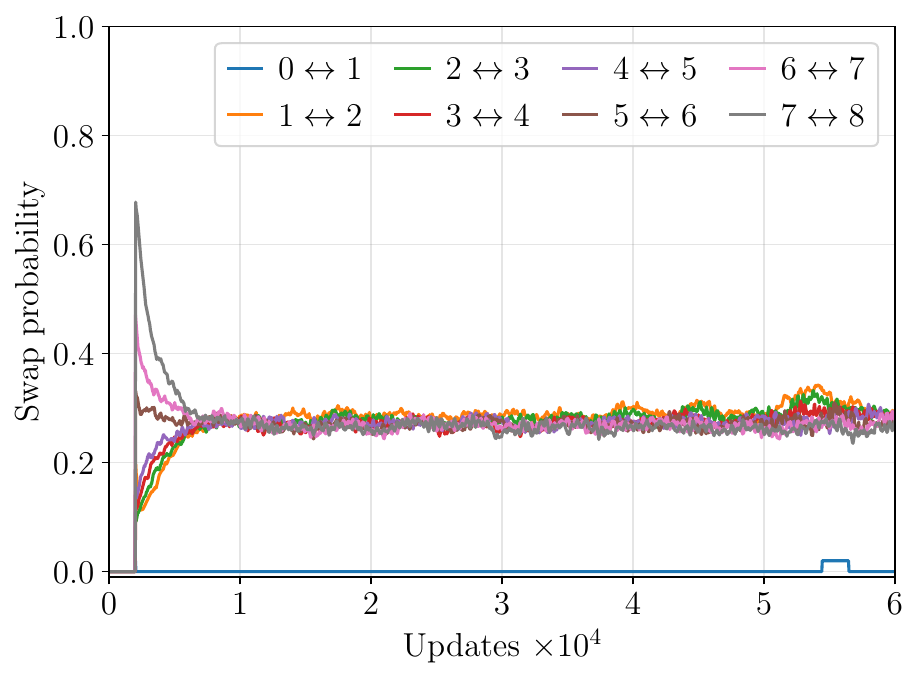} \label{fig:precipice-swap-probs-with}}
  \caption{Running average of the swap probability with 10 replicas for a single simulation (a) without temperature optimization and (b) with temperature optimization  for the precipice problem.  The running average is performed over 50 swap attempts (or equivalently 200 SR updates).}
  \label{fig:precipice-swap-probs}
\end{figure}

To demonstrate that the temperature optimization performs as expected, we show in Fig.~\ref{fig:precipice-swap-probs} a running average of the swap probability calculated over 50 swap attempts.  In Fig.~\ref{fig:precipice-swap-probs-without}, we see that in the case of no temperature optimization and with our choice of initial temperature distribution, there are pairs of replicas, specifically replica pairs (2,3) and (3,4), that exhibit effectively zero swap probability.  This prevents configurations trained at higher temperatures to be exchanged with those at lower temperatures, and we do not observe a single case of finding the ground state with 50 independent simulations.

\begin{figure}
    \centering
    \subfigure[]{\includegraphics[width=2.95in]{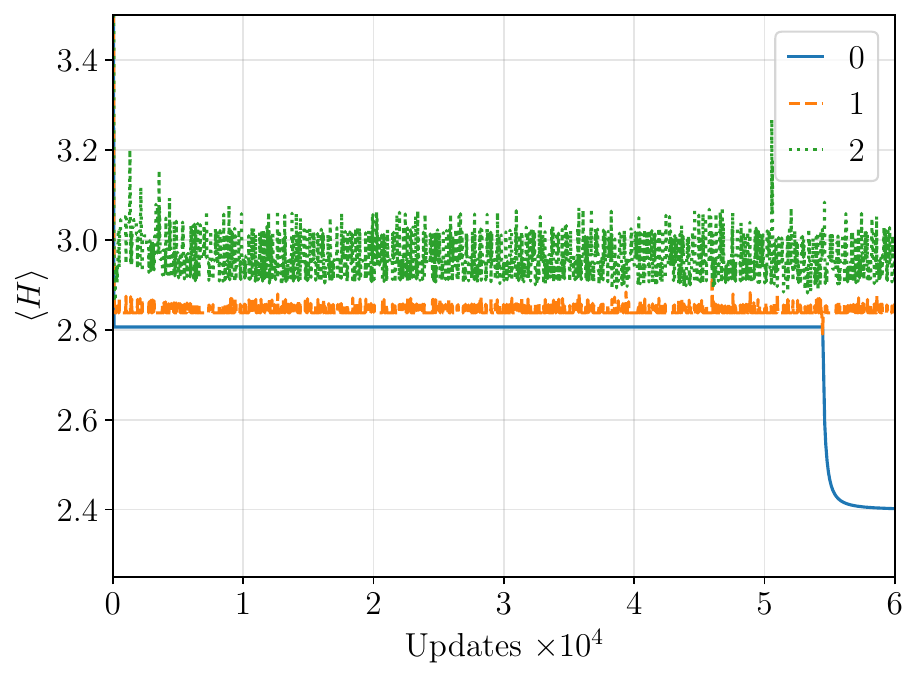} \label{fig:precipice-energy}}
  \subfigure[]{\includegraphics[width=2.95in]{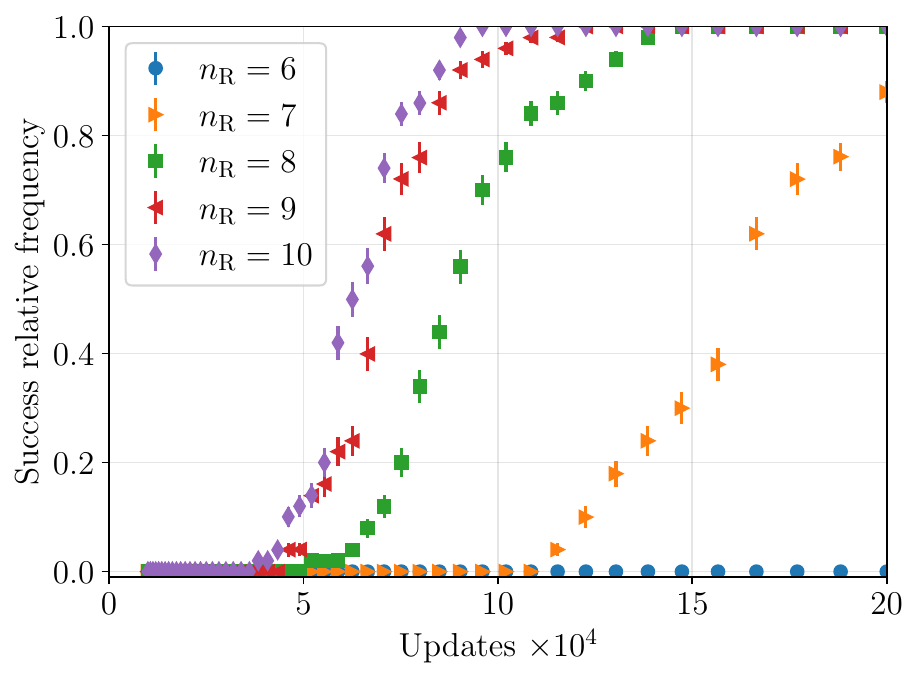}   \label{fig:precipice-CDF}}
  \caption{(a) Expectation value of the precipice Hamiltonian (Eqt.~\eqref{eqt:PrecipiceH}) for the lowest three temperature replicas during training with $n_{\mathrm{R}} = 10$ corresponding to the simulation in Fig.~\ref{fig:precipice-swap-probs-with}, which includes temperature optimization. (b) Frequency of finding the ground state with temperature optimization for the precipice problem.  The results are calculated for 50 independent simulations, with the error bars corresponding to the $2 \sigma$ confidence interval calculated using a bootstrap.}
  \label{fig:precipice-success}
\end{figure}

In contrast, in Fig.~\ref{fig:precipice-swap-probs-with} we show how the swap probability for all pairs of replicas converges to a nonzero value.  The exact nonzero value depends on the number of replicas used: a higher value is reached as more replicas are used. Unlike the case without temperature optimization, the algorithm is now able to find the ground state. For the particular case illustrated in Fig.~\ref{fig:precipice-swap-probs-with}, this coincides with a `blip' in the swap probability between the 0 and 1 replicas, where the configuration at $T_{\min}$ finds a lower energy and its configuration is swapped with the configuration at $T=0$. This can be seen in Fig.~\ref{fig:precipice-energy} where the expectation value of the Hamiltonian for the zero temperature replica is shown to finally reach the ground state at that time.

In Fig.~\ref{fig:precipice-CDF}, we show how the frequency of finding the ground state depends on the number of replicas. For 6 replicas and fewer, the swap probabilities are too low to allow the algorithm to find the ground state with the fixed number of updates we performed.  As the number of replicas is increased, the number of updates needed to find the ground state is reduced, but the benefit of having more replicas diminishes once the swap probability becomes too large.  This is consistent with what is observed in more `traditional' studies using PT where an optimal swap probability is observed \cite{Kone2005}.
\subsection{$J_1$-$J_2$ Model with RBMs} \label{sec:J1J2RBM}
%
As a second test of our method, we study the $J_1$-$J_2$ model on a square grid with open boundary conditions. The Hamiltonian of the system is given by:
\beq
 \hat{H} = J_1 \sum_{\langle i,j \rangle} \vec{\sigma}_i \cdot \vec{\sigma}_j + J_2 \sum_{\langle \langle i,j \rangle \rangle} \vec{\sigma}_i \cdot \vec{\sigma}_j 
\eeq
where $\vec{\sigma}_i = \hat{\sigma}_i^x \hat{x} + \hat{\sigma}_i^y \hat{y}+ \hat{\sigma}_i^z \hat{z}$ is the vector of Pauli operators acting on site $i$, $\langle i,j \rangle$ denotes nearest neighbor interactions, and $\langle \langle i,j \rangle \rangle$ denotes next nearest neighbor interactions. Because the Hamiltonian commutes with the total $z$-magnetization operator $\hat{M}^z = \sum_{i=1}^n \hat{\sigma}_i^z$, the energy eigenstates can be expressed as simultaneous eigenstates of $\hat{M}^z$. For simplicity, we assume that the number of sites is even, in which case the ground state has is an eigenstate of $\hat{M}^z$ with eigenvalue 0. This means we can restrict the Monte Carlo sampling in SR to computational basis states with Hamming weight $n/2$ configurations (or equivalently to spin configurations with an equal number of up and down spins).

For our first case, we focus on a $4 \times 5$ system with $J_1 =1$ and $J_2 = 0.695$, training with an RBM. This is close to the point where the gap between the first excited state and ground state is smallest, near $J_2=0.7$, and where we find the greatest difficulty in training an RBM.

\begin{figure}[t] 
  \centering
  \subfigure{\includegraphics[width=2.95in]{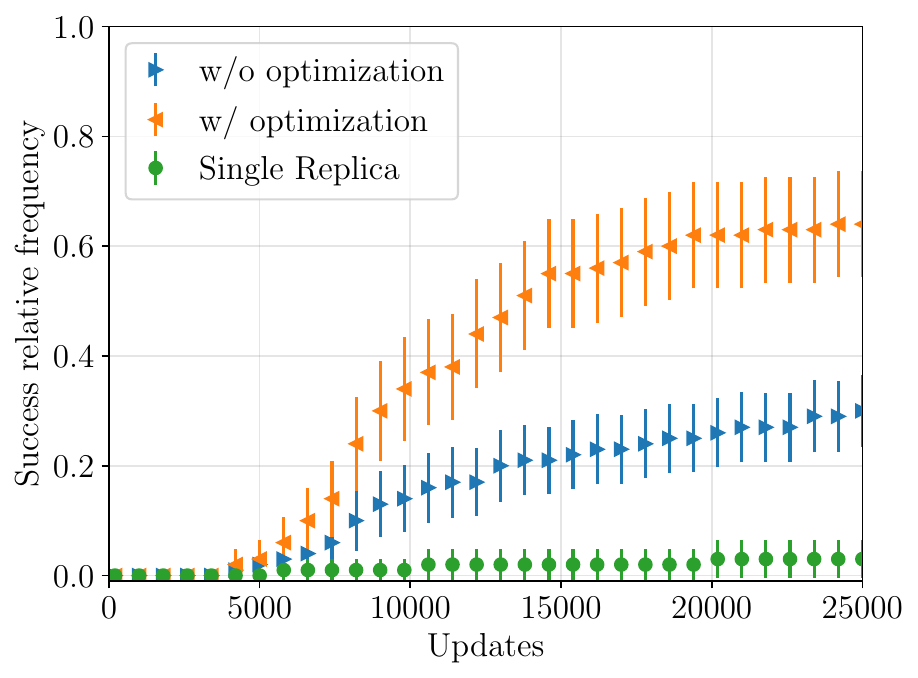}\label{fig:j1j2-CDFA}} 
  
  \subfigure{\includegraphics[width=2.95in]{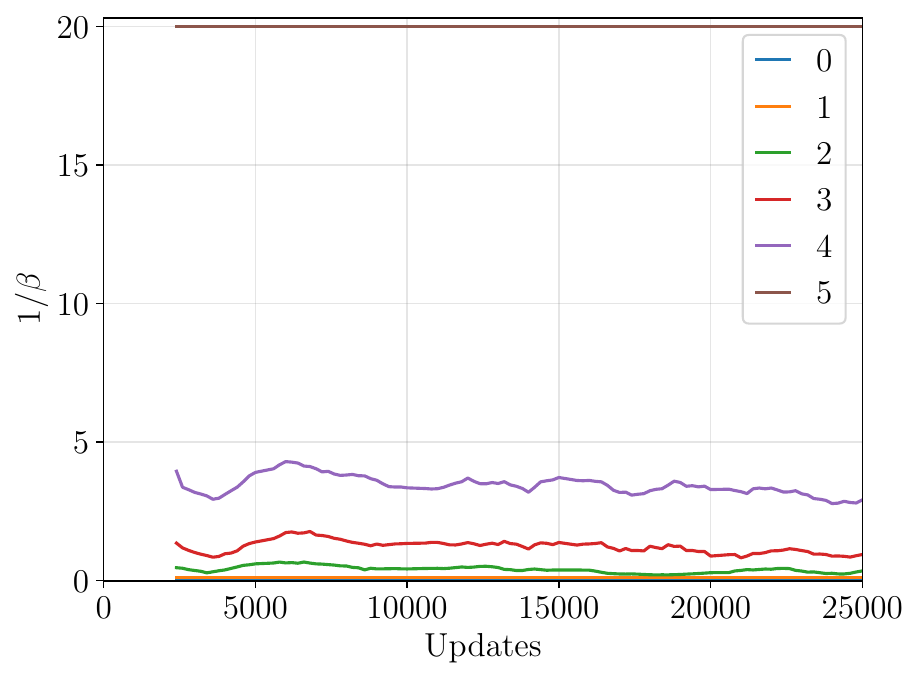}\label{fig:j1j2-CDFB}} 
  
  \caption{ (a) Relative frequency of finding an energy below the first excited state ($-34.70212$) of the open boundary condition 4$\times$5 $J_1$-$J_2$ problem at $J_2=0.695$ with increasing number of updates in PT for 100 independent simulations, with the error bars corresponding to the $2 \sigma$ confidence interval calculated using twice the standard error of the mean. For the PT simulations ($\triangleleft$ and $\triangleright$ data points) we use $n_{\mathrm{R}}=6$ replicas. (b) The running average value of $\beta$ as a function of updates for a representative run when the temperature is allowed to adjust.}
  
\end{figure}
We use an RBM ansatz with $n=20$ visible units and $m=40$ hidden units for our variational wavefunction. This ansatz is likely not expressive enough to fully capture the ground state; we have not observed energies reaching the ground state in any of our simulations. However, it is expressive enough to capture states with energies below the first excited state, which we term as success throughout the rest of this section.

The initial temperature distribution is picked to be $ T_k = T_{\min} + (T_{\max} - T_{\min})\left(\frac{k-1}{n_{\mathrm{R}}-2} \right)^{3}$, where $k = 1, \dots, n_{\mathrm{R}}-1$. We initialize with the parameter choices $T_{\min} = 0.1$ and $T_{\max} = 20$. We give further details of our simulation settings in Appendix~\ref{app:J1J2Settings}. 

In order to quantify the advantage of our method, we use the likelihood of the simulation reporting an energy beneath the first excited state as opposed to the mean energy achieved. This is because we are not trying to gauge the expressibility of the wavefunction but the computational efficiency of PT. In Fig.~\ref{fig:j1j2-CDFA} we measure this likelihood and compare the case of a single replica to the case of PT with 6 replicas. In the case of PT, we consider the two cases of allowing temperature optimization throughout the training and holding the temperature fixed throughout the training. The $J_1$-$J_2$ problem is inherently difficult for a single replica RBM system to solve without any PT, with a success rate of only 3\% $\pm$ 3\% after $2.5 \times 10^4$ updates. PT without any temperature optimization improves the overall success rate to 30\% $\pm$ 9\%, demonstrating a clear improvement. Once we allow for temperature optimization, we achieve a significantly higher success rate of 65\% $\pm$ 9\%.
In Fig.~\ref{fig:j1j2-CDFB}, we show how the temperature $1/\beta$ changes over time for a typical run where we allow the temperatures to optimize.
We find that our initial polynomial distribution of temperatures is relatively good and only small adjustments are made throughout the run. 
This demonstrates that optimizing temperatures can increase success probability even for systems that are initially well tuned.

\begin{figure*}
    \centering
  \subfigure[]{\includegraphics[width=2.95in]{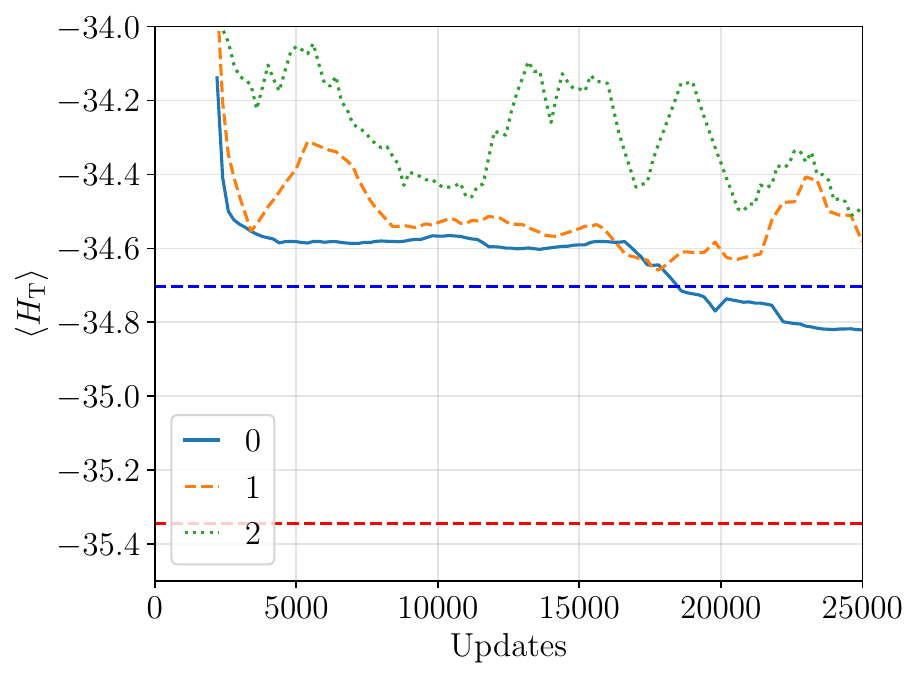} \label{fig:j1j2-swap-probsA}}
  \subfigure[]{\includegraphics[width=2.95in]{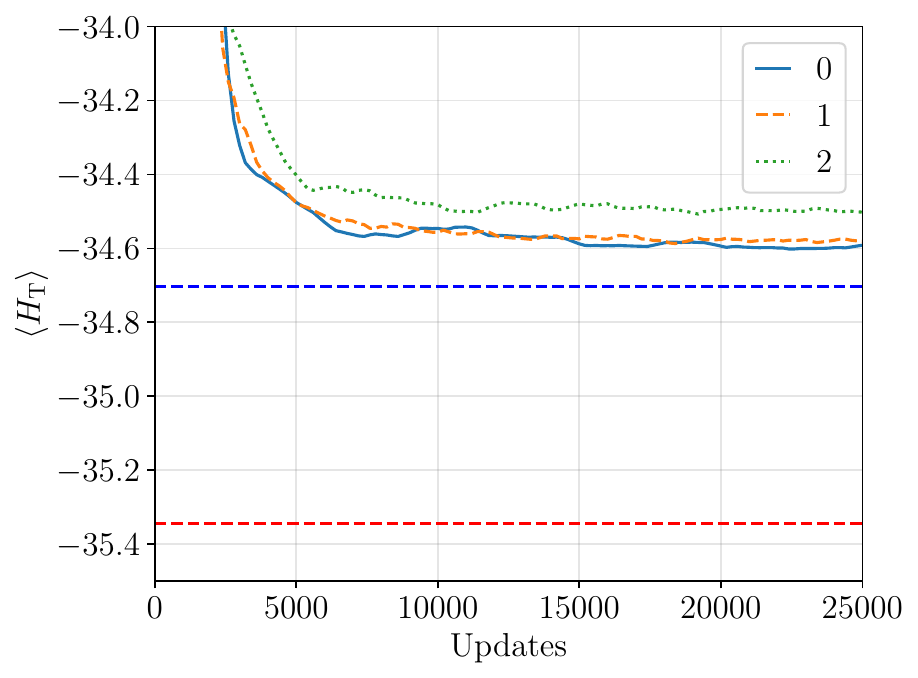}\label{fig:j1j2-swap-probsB}}
  \subfigure[]{\includegraphics[width=2.95in]{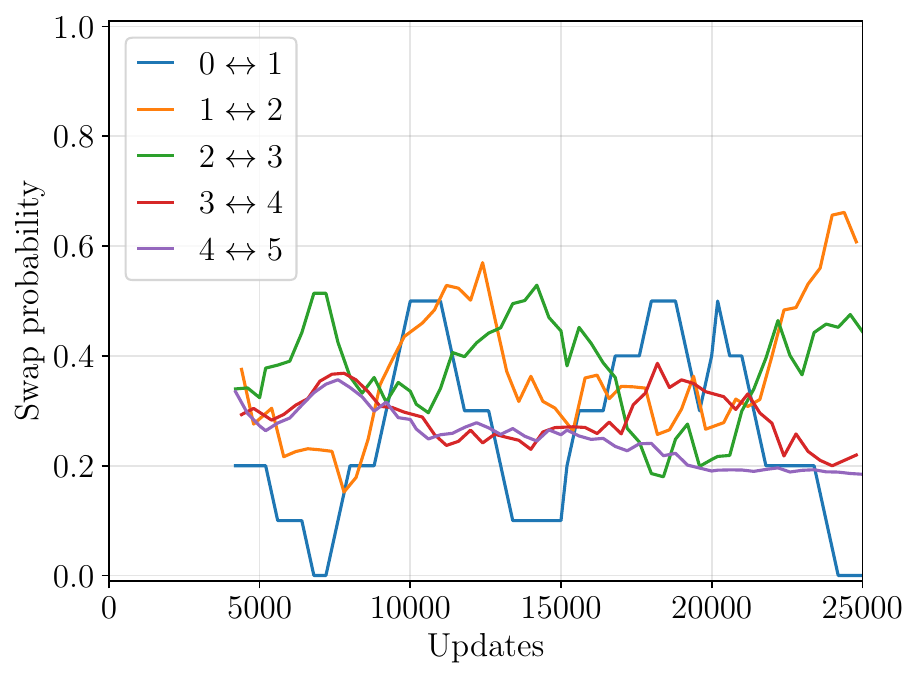} \label{fig:j1j2-swap-probsC}}
  \subfigure[]{\includegraphics[width=2.95in]{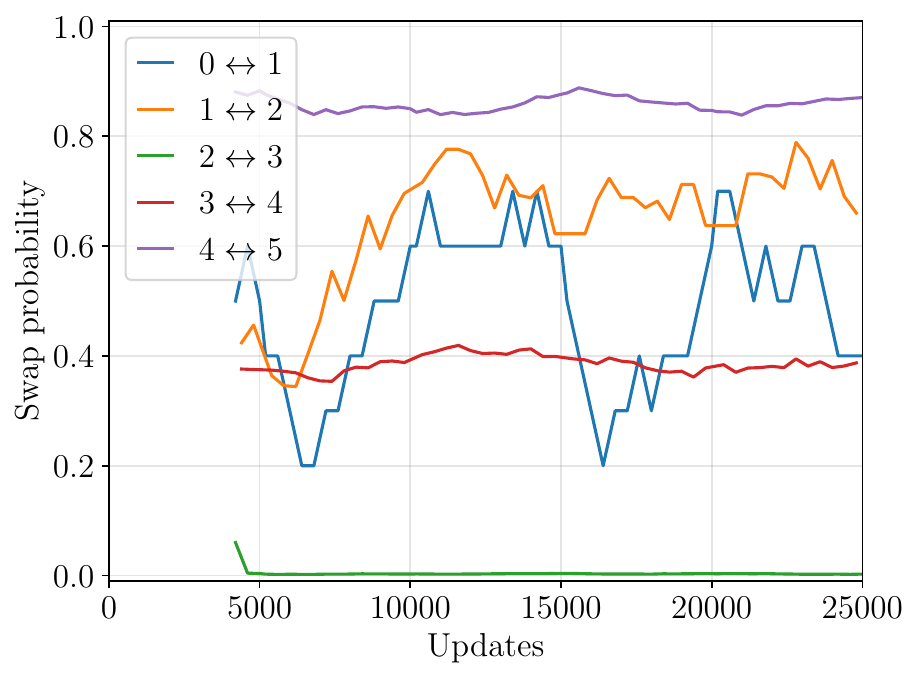}\label{fig:j1j2-swap-probsD}}
  \caption{The running average of the energy of the first 3 replicas for a representative run of the open boundary conditions 4$\times$5 $J_1$-$J_2$ model where the temperatures are (a) allowed to adjust and (b) held fixed.
  The red dashed line corresponds to the ground state energy ($-35.34479$) and the blue dashed line to the first excited state energy ($-34.70212$). The running average of the probability of swapping between different replicas for the corresponding run for the case of (c) allowing the temperature to adjust and (d) holding the temperatures fixed.}
  \label{fig:j1j2-swap-probs}
\end{figure*}
The reason for this improvement can be seen by examining the swap probabilities between each of the different replicas for a typical run, as shown in Fig.~\ref{fig:j1j2-swap-probs}.
When the temperatures are allowed to optimize, the swap probabilities converge near 0.3, although the values are still noisy due to the limited sample size and number of updates.
For transferring into the zero replica (which contains the pure target Hamiltonian), the probability of swapping is a binary 0 or 1, based on whether replica 1 has a lower target energy.
When we take the running average in Fig.~\ref{fig:j1j2-swap-probsC}, this corresponds to the averaged probability moving up for update regions where there is more transfer into the zero replica.
This move toward high probabilities for swapping into/out of the zero replica is particularly pronounced at $10^4$ updates and $1.6 \times 10^4$ updates.
The latter of these update states corresponds precisely with when the energy of the zero replica begins to lower, eventually dipping below the first excited state.
In contrast, when the temperatures are not allowed to optimize as in Fig.~\ref{fig:j1j2-swap-probsD}, the probabilities of swapping never converge toward any value. 
Furthermore, the probability of swapping between the 2nd and 3rd replicas remains at essentially 0 throughout the entirety of the simulation. 
This leads to an inability for any configurations from replicas 3 to 5 to ever make their way to the 0 replica, which likely leads to the simulation never reaching energies below the first excited state. 
While these results are for two specific runs of PT for this problem, they are representative of how the algorithm achieves success for a majority of our simulations. 
This shows that optimizing temperatures allows the PT algorithm to perform better by ensuring that states are mixed in from the higher temperatures in the distribution at a steady rate. 

\begin{figure} 
  \centering
  \subfigure[]{\includegraphics[width=2.95in]{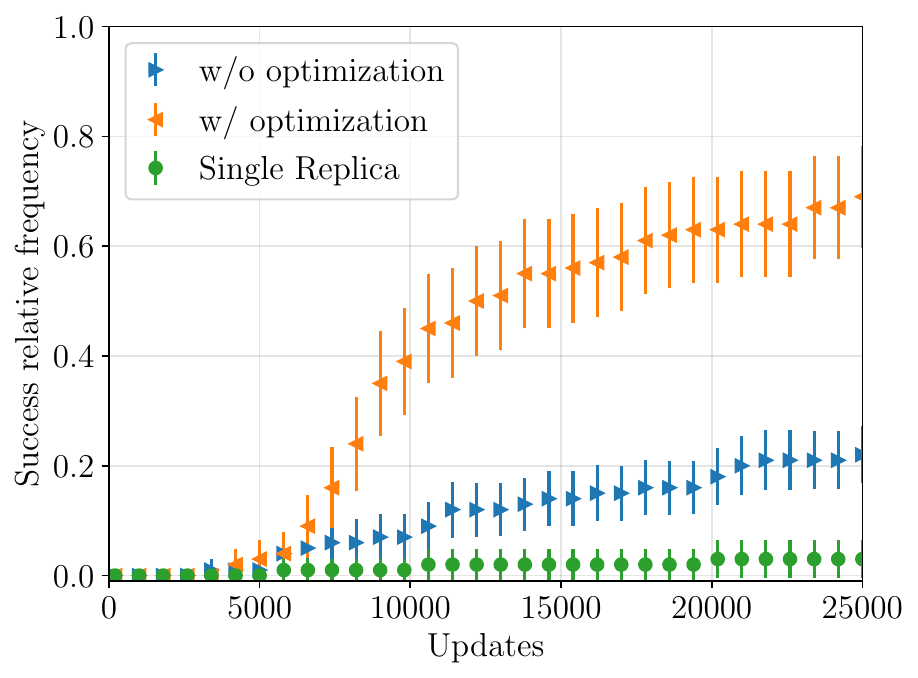}\label{fig:j1j2-linearA}}
  \subfigure[]{\includegraphics[width=2.95in]{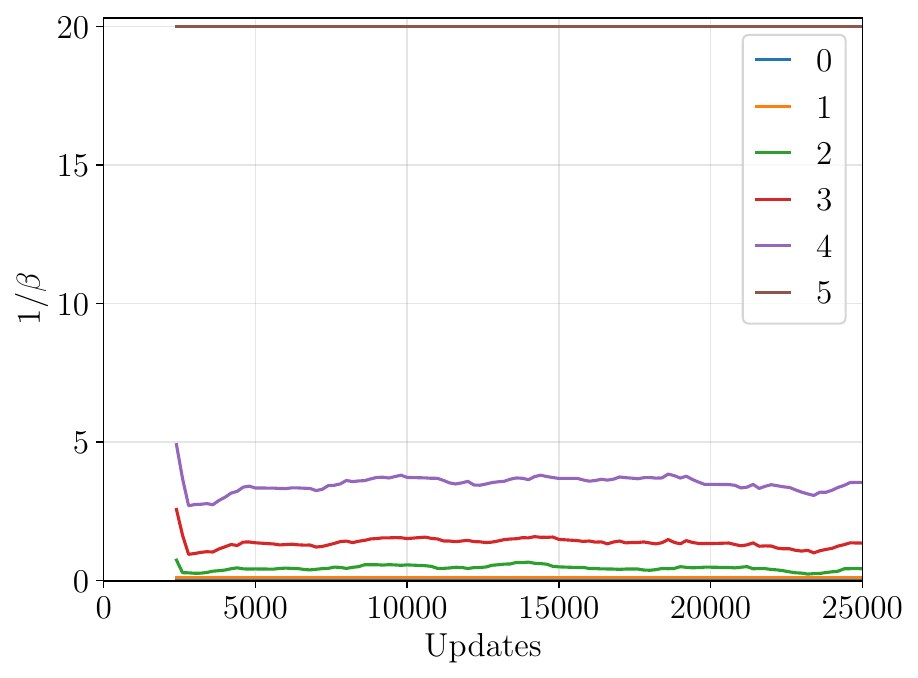}\label{fig:j1j2-linearB}}
  \caption{(a) Relative frequency of finding an energy below the first excited state ($-34.70212$) of the open boundary condition 4$\times$5 $J_1$-$J_2$ problem at $J_2=0.695$ when starting from an initial linear temperature distribution instead of a polynomial distribution over 100 independent runs, with the error bars corresponding to the $2 \sigma$ confidence interval calculated using twice the standard error of the mean. (b) The running average value of $\beta$ as a function of updates for a representative run when the temperature is allowed to adjust.
  }
  \label{fig:j1j2-linear}
\end{figure}

We further find that the temperature optimization algorithm is relatively robust to the initial choice of temperature distribution. In Fig.~\ref{fig:j1j2-linear}, we explore a linear initial distribution of temperatures, $T_k = T_{\min} + (T_{\max} - T_{\min})\left(\frac{k-1}{n_{\mathrm{R}}-2} \right)$, where $k = 1, \dots, n_{\mathrm{R}}-1$. The probability of success at $J_2=0.695$ nearly exactly matches what we see for the polynomial distribution in Fig.~\ref{fig:j1j2-CDFA}. Without optimizing the temperatures, we see a success rate of $0.22 \pm 0.082$. Once we allow the temperatures to optimize, the success rate improves to $0.68 \pm 0.093$. In Fig.~\ref{fig:j1j2-linearB}, the $\beta$ values in a representative example run adjust significantly away from their initial linear distribution within just approximately $10^3$ updates. This distribution of temperatures closely matches the distribution that we see when starting with an initially polynomial distribution. From these results, we conclude that our algorithm is robust to choices of the initial temperature distributions.

\begin{figure}
  \centering
  \includegraphics[width=2.95in]{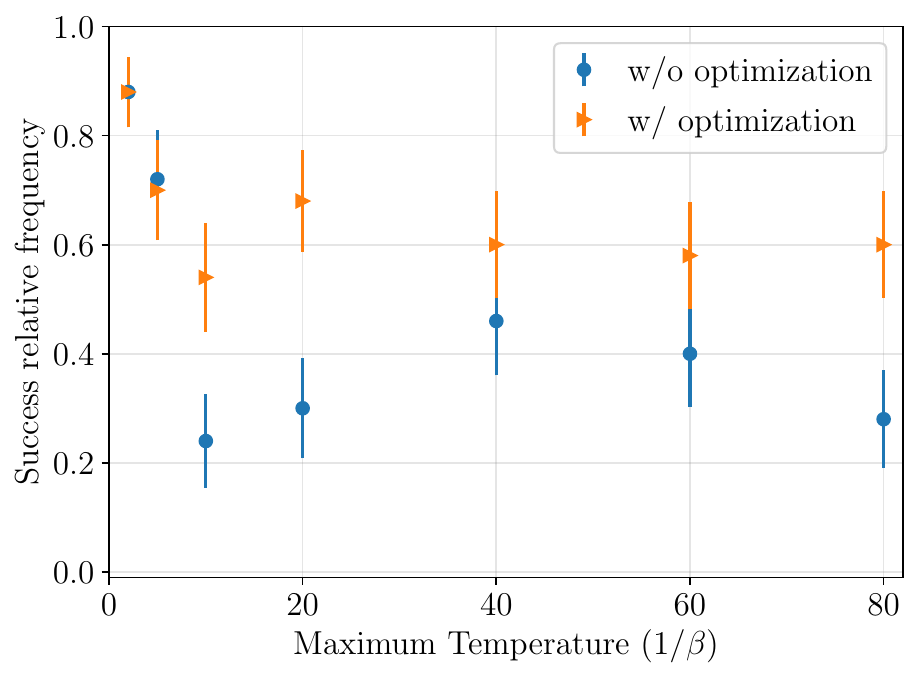} 
  \caption{Relative frequency of finding an energy below the first excited state of the open boundary condition $4\times5$ $J_1$-$J_2$ problem as a function of the maximum temperature.}
  \label{fig:j1j2-maxT}
\end{figure}

We do, however, observe a relatively strong dependence on the choice of minimum and maximum temperatures, which are held fixed throughout the simulation.
In Fig.~\ref{fig:j1j2-maxT}, we vary the maximum temperature $T_{\max}$ for a system with $n_R=6$ and an initial polynomial distribution of temperatures and note how the success rate changes. We see that for maximum temperatures greater $\geq 10$, the success rate for simulations where the temperature distribution is  optimized holds steady at $\approx 0.60$. For simulations without temperature optimization, the success rate varies from $\approx 0.20$ to $0.40$. Since  setting the maximum temperature also affects the initial temperature distribution, this variation is expected. 

However, as the maximum temperature decreases below 10, the success rate for the simulations drastically increases, with a choice of $T_{\max}=2$ giving a success rate of $0.88$ for both simulations with and without temperature optimization. This shows that the choice of $T_{\max}=20$ explored in the earlier figures was not optimal for maximizing success. Therefore, the success frequency of our algorithm is still highly sensitive to the initial choice of the maximum temperature. Nevertheless, the temperature optimization does provide robust improvement to the success rate even for suboptimal choices of initial temperature, which can speed up the process of finding viable solutions. Future work will focus on automating the selection of the $T_{\max}$ parameter. 

In Appendix~\ref{app:hyperparameter}, we explore the dependence of our algorithm's performance on the choice of additional hyperparameters such as the number of replicas and the number of samples. We find that the performance can depend strongly on these values. Nonetheless, in all cases tested, allowing the temperatures to optimize within the simulation provides at least as much success as systems without optimization, and typically substantially improves success rates for minimal additional computational cost. 
\subsection{$J_1$-$J_2$ Model with Feedforward Networks}
\label{sect:FFN}
%
\begin{figure}[!t]
    \centering
    {\includegraphics[width=2.95in]{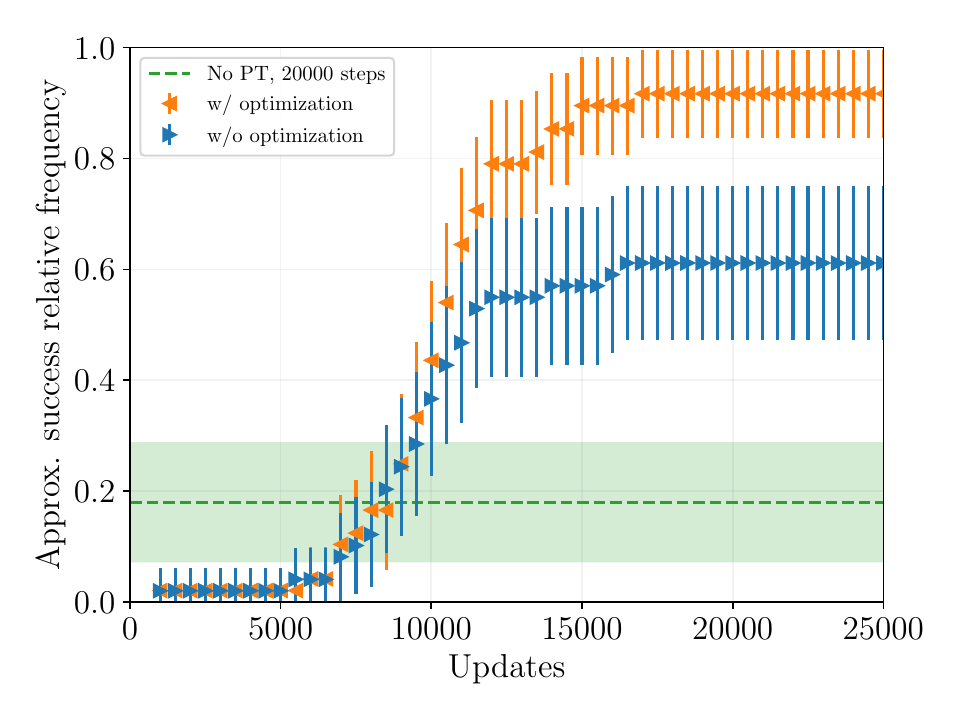}}\label{fig:FFNb}
    \caption{Success relative frequency on the $6\times 6$ $J_1$-$J_2$ model with $J_2=0.62$ using a feedforward network with $(48,24,24)$ hidden units over 50 independent runs. The green dashed line corresponds to the success frequency of independent single-replica runs (without PT) after $2 \times 10^4$ updates with the shaded region corresponding to the $2\sigma$ confidence interval calculated using a bootstrap. The blue triangles correspond to PT with 5 replicas without temperature optimization, and the orange triangles correspond to PT with 5 replicas with temperature optimization. Success is defined as finding a state with an average energy that is at least one standard error below the first excited of $-70.07405$. The error bars correspond to the $2\sigma$ confidence interval calculated using a bootstrap.}
    \label{fig:feedforwardresults}
\end{figure}

As a final test, we study the $J_1$-$J_2$ on a square lattice with periodic boundary conditions for a $6 \times 6$ system. In this case, we choose $J_1 = 1$ and $J_2 = 0.62$. This is close to the point where the energy gap between the first excited state and the ground state is smallest, near $J_2=0.58$. For this study, we use a three-layer feedforward neural network instead of an RBM, and we denote the number of hidden nodes in each layer by $(h_1,h_2,h_3)$. This network is more expressive than the RBMs used in Sec.~\ref{sec:J1J2RBM} and is intended to test our method on a more expressive architecture. For this system size, exact sampling is computationally prohibitive and prevents us from evaluating the energy of the NQS exactly, so we employ a different criterion to define success. We take a running mean of the energy and standard deviation over 50 steps and if the average energy reported is beneath $-70.07405$ (computed using DMRG~\cite{White1992,White1993,Schollwock2005,Schollwock2011} with a bond dimension of 2048), by more than a third of the mean standard deviation, we freeze the parameters and sample at that step for $10^3$ steps with 1024 samples each. Each of these sampling runs gives a sample mean energy and a sample standard deviation, from which we can calculate a mean of the means energy and a standard error of the mean.  If the mean of the mean energies is one standard error below the first excited state energy, then we flag the simulation as a possible success. 
When the simulation terminates we again sample, this time for $10^3$ steps with 512 samples each in order to confirm the success of the simulation. We find that all simulations that are flagged successful at intermediate points in the simulation are confirmed successful with this final test.

For our three-layer feedforward network, we choose a network structure such that the first hidden layer has twice the hidden units as the next two hidden layers, and we choose $(h_1, h_2, h_3) = (48,24,24)$.  We find this is expressive enough to yield states with energies below the first excited state.

For the PT runs, we use a linear initial temperature distribution: $T_k = T_{\min} + (T_{\max} - T_{\min})\left ( \frac{k-1}{n_R-2} \right)$. We do this to highlight the adaptivity of the temperature optimization scheme. We start performing replica swaps after the $10^3$-th SR update step when energies have had time to converge, after which we perform $10^2$ SR update steps between replica swaps and temperature updates. Some of the implementation details of the variational algorithm used for the feedforward network are different from those of Sec.~\ref{sec:J1J2RBM}, and we give these details in Appendix~\ref{app:FeedForward}.

We show the success of PT using the $(48,24,24)$ feedforward network in Fig.~\ref{fig:feedforwardresults}. We include the mean energies of these simulations in Appendix~\ref{app:MeanEnergies}, but we again find success probability to be a more informative metric. We observe a clear advantage to using PT over not using PT in terms of success frequency, and we also begin to see a separation between the simulations with and without temperature optimization after approximately $10^4$ update steps. After $1.5\times10^4$ steps, the increase in success rate for the case without temperature optimization slows down and possibly saturates near $60\%$, suggesting that the system can get stuck in local minima. This is in contrast to the case with temperature optimization  where a properly tuned temperature distribution can help it escape from such local minima and succeed with a rate near $90\%$.

\section{Conclusions} \label{sec:conclusions}

In this work, we demonstrated the impact of the replica temperature distribution on the effectiveness of PT to help train an ANN to represent a low-energy quantum state. Training with PT already shows increased success rates over training with a single replica, and by adaptively adjusting the temperature distribution such that all pairs of replicas swap with equal probability, we reach even higher success frequencies at quicker rates with negligible computational cost. We attribute this improvement to better escape rates from local minima. These results are more pronounced for problems where the training is difficult.  This can arise from deep local minima in the energy potential, which was illustrated by the Precipice problem, but also when the number of parameters gets larger for more expressive networks, as was demonstrated by our different network sizes for the $6 \times 6$ $J_1$-$J_2$ model. We expect that the difficulties encountered for the $J_1$-$J_2$ problem will only be exacerbated at larger problem sizes, so we can expect temperature optimization to be crucial in that case. 

We have focused on neural networks that are generic to demonstrate the effectiveness of our proposed methodology. Replacing the RBM or feedforward networks with more state-of-the-art networks adopted for NQS such as GCNN's~\cite{Roth2021,Roth2023} is straightforward. While it remains to be tested, we expect PT to be useful with larger architectures due to the difficulty inherent in their training. The FermiNet architecture~\cite{spencer2020}, for example, typically utilizes nearly $\mathcal{O}(10^6)$ parameters and $\mathcal{O}(10^5)$ training iterations alongside a pretraining procedure. PT may enable either the number of parameters or total training iterations to be decreased for larger networks like this. 
While larger networks make finding local minima close to the global minimum of the training landscape easier, they also make finding the global minimum exponentially more difficult~\cite{Choromaska2014TheLS}. If all of the local minima are a good approximation of the ground state, then increasing the network size alone would be a successful strategy. However, it is also possible that the global minimum is a qualitatively different state from any of the local minima, in which case finding any local minimum may not be sufficient.
We therefore believe PT to be particularly useful in cases where finding the global minimum with higher efficiency is needed. When PT is used in tandem with large networks, we expect it to increase the probability of converging to the global minimum or to a minimum close to the limit of the network's expressibility. It may also decrease the number of training iterations required to converge and allow us to use smaller networks to achieve similar energies. All three of these advantages in efficiency become more important with larger networks.

An important feature of the approach we advocate is that it introduces minimal additional computational cost to the PT algorithm, and in all the cases we have studied, optimizing the temperature distribution improves the performance of the PT algorithm or at least provides equal performance. It introduces only two additional hyperparameters, the minimum and maximum temperature of the distribution, and we leave automating these hyperparameters for future work.

\begin{acknowledgements}
We would like to thank the UNM Center for Advanced Research Computing, supported in part by the National Science Foundation, for providing the high performance computing resources used in this work.

This material is based upon work supported by the National Science Foundation under grant number 2037755. This material is based upon work supported by the Air Force Office of Scientific Research under award number FA9550-22-1-0498. Any opinions, findings, and conclusions or recommendations expressed in this material are those of the author(s) and do not necessarily reflect the views of the United States Air Force.

This work was performed, in part, at the Center for Integrated Nanotechnologies, an Office of Science User Facility operated for the U.S. Department of Energy (DOE) Office of Science. 

The Flatiron Institute is a division of the Simons Foundation.
 
Sandia National Laboratories is a multi-mission laboratory managed and operated by National Technology \& Engineering Solutions of Sandia, LLC (NTESS), a wholly owned subsidiary of Honeywell International Inc., for the U.S. Department of Energy’s National Nuclear Security Administration (DOE/NNSA) under contract DE-NA0003525. This written work is authored by an employee of NTESS. The employee, not NTESS, owns the right, title and interest in and to the written work and is responsible for its contents. Any subjective views or opinions that might be expressed in the written work do not necessarily represent the views of the U.S. Government. The publisher acknowledges that the U.S. Government retains a non-exclusive, paid-up, irrevocable, world-wide license to publish or reproduce the published form of this written work or allow others to do so, for U.S. Government purposes. The DOE will provide public access to results of federally sponsored research in accordance with the DOE Public Access Plan \url{https://www.energy.gov/downloads/doe-public-access-plan.}

\end{acknowledgements}

%

\appendix
\section{Simulation settings for precipice problem} \label{app:PrecipiceSettings}
%
For the simulations on the Precipice problem discussed in Sec.~\ref{sec:Precipice}, we use SR with an adaptive learning rate.  This requires solving the linear system of equations in Eqt.~\eqref{eqt:SRupdate}.  To do this, we first perform an explicit regularization of $s$ by replacing its diagonal elements by \cite{Carleo2017}:
\beq
s_{kk} \to s_{kk}(1 + \lambda(p))
\eeq
where $p$ is the current update step and we choose $\lambda(p) = \max(\lambda_0 b^p, \lambda_{\min}$ with $\lambda_0 = 100, b=0.9, \lambda_{\min} = 10^{-4}$. We then use MINRES-QLP \cite{Choi2014} to solve the equations, using a tolerance of $10^{-6}$ and a maximum of $10^3$ iterations.  The total number of SR updates in the simulation is $2 \times 10^5$.

The learning rate $\eta$ in Eqt.~\eqref{eqt:SRupdate} is adjusted adaptively using a Heun second order consistent integrator \cite{Schmitt2020} with an error of $10^{-8}$.  
For PT, the lowest finite temperature in the simulation is given by $T_1 = 0.05$ and the largest temperature is given by $T_{n_{\mathrm{R}}-1} = 10$. PT swaps are performed at every $n_{\mathrm{swap}} = 20$ updates, alternating between odd and even neighbors as discussed in the main text.  The initial temperature distribution is picked to be $ T_k = T_{\min} + (T_{\max} - T_{\min})\left(\frac{k-1}{n_{\mathrm{R}}-2} \right)^{3}$, where $k = 1, \dots, n_{\mathrm{R}}-1$. In the case of optimizing the temperature distribution, the update of the temperature distribution occurs every 200 updates.
\section{Simulation settings for the $J_1$-$J_2$ problem} 
\subsection{RBM settings} \label{app:J1J2Settings}
For the simulations for the $4 \times 5$ $J_1$-$J_2$ problem using the RBM shown in the main text, we use a similar procedure as discussed in Appendix~\ref{app:PrecipiceSettings}. We use MINRES-QLP with the same tolerance and maximum number of updates. We use the same SR with an adaptive learning rate implementation but now with an allowed error of $10^{-5}$. 

Because the ground state is in the zero-magnetization sector, we restrict our Monte Carlo sampling to computational basis states with fixed Hamming weight equal to 10.  We take 200 samples per SR update step, which are used to estimate the quantities $f$ and $s$ in Eq.~\eqref{eqt:sandf}. We also use these samples to calculate the running average of the energy used in calculating the parallel tempering probabilities (Eq.~\eqref{eqt:SwapProbability}) as well as the temperature updates~(Eq.~\eqref{eqt:temperatureUpdate}). For the plots of the energy in Fig~\ref{fig:j1j2-swap-probs}, we calculate the energy of each replica exactly by sampling all Hamming weight 10 states, but the exact energy is not used in the training in any way.
When optimizing the temperatures, we start with 400 updates of ``burn in'' after which we update the temperature distribution every 200 updates.
PT swaps are performed at every $n_{\mathrm{swap}} = 100$ updates, alternating between odd and even neighbors as discussed in the main text. 
For all these simulations, we use a total of 25000 updates.

We discuss the sensitivity of our results to some of the parameter choices in Appendix~\ref{app:hyperparameter}.

\subsection{Feedforward Network Settings} \label{app:FeedForward}
The results for the feedforward neural networks in Sec.~\ref{sect:FFN} were obtained using the python package NetKet \cite{netket3:2022}. This package can handle arbitrary neural network architectures and uses autodifferentiation implemented with JAX \cite{jax2018github}. We have used NetKet's default version of SR, which uses the conjugate gradient method to perform the inversion in Eqt.~\eqref{eqt:SRupdate}.

In contrast to the RBM simulations which use a Heun second order consistent integrator as well as SR to provide accurate gradients, we have used SR alone here. We expect an adaptive learning rate would be useful on this larger system, but it is more computationally expensive. We choose a learning rate of $10^{-3}$ and train for $10^4$ total steps in the case of the largest feedforward network and $2.5 \times 10^4$ steps for the two smaller feedforward networks. Parallel tempering updates do not begin until step $10^3$ to allow the optimization time to converge. After this, we swap replicas and perform temperature optimization (if indicated) every $10^2$ steps.

\section{Hyperparameter dependence in the $J_1$-$J_2$ problem} \label{app:hyperparameter}
\begin{figure} [ht]
  \centering
  \includegraphics[width=2.95in]{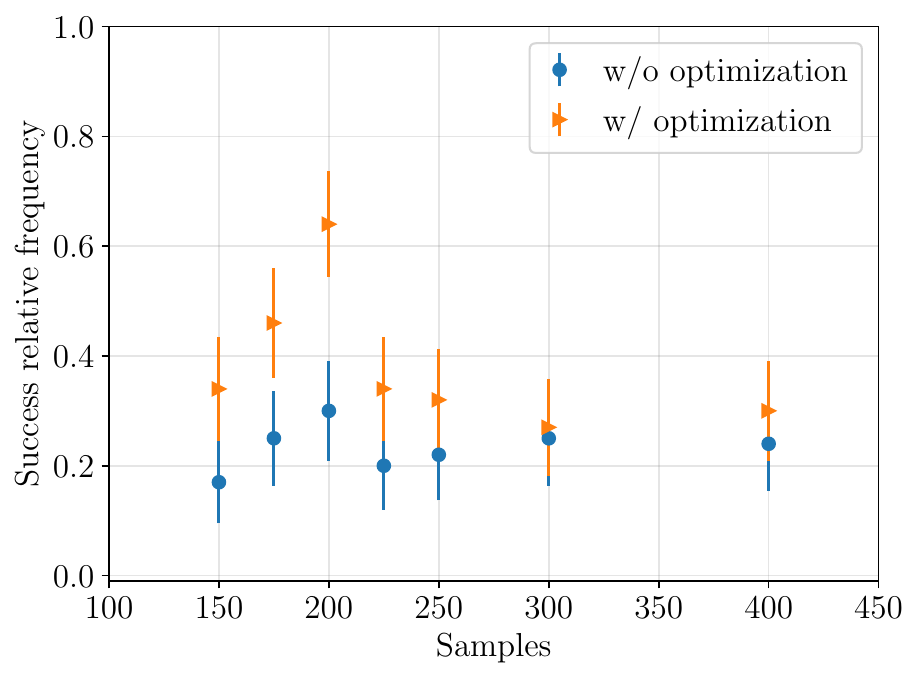} 
  \caption{Relative frequency of finding an energy below the first excited state of the open boundary condition 4$\times$5 $J_1$-$J_2$ problem when varying the number of samples taken over 100 independent runs, with the error bars corresponding to the $2 \sigma$ confidence interval calculated using twice the standard error of the mean.}
  \label{fig:j1j2-samples}
\end{figure}

\begin{figure}[htb]
  \centering
  \subfigure[]{\includegraphics[width=2.95in]{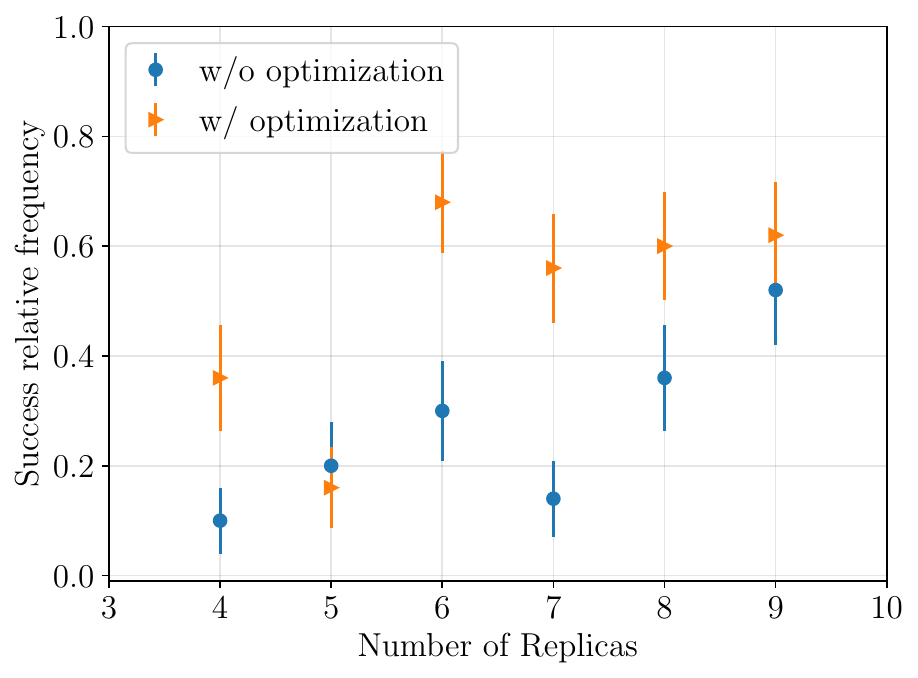} \label{fig:j1j2-replicasA}}
  \subfigure[]{\includegraphics[width=2.95in]{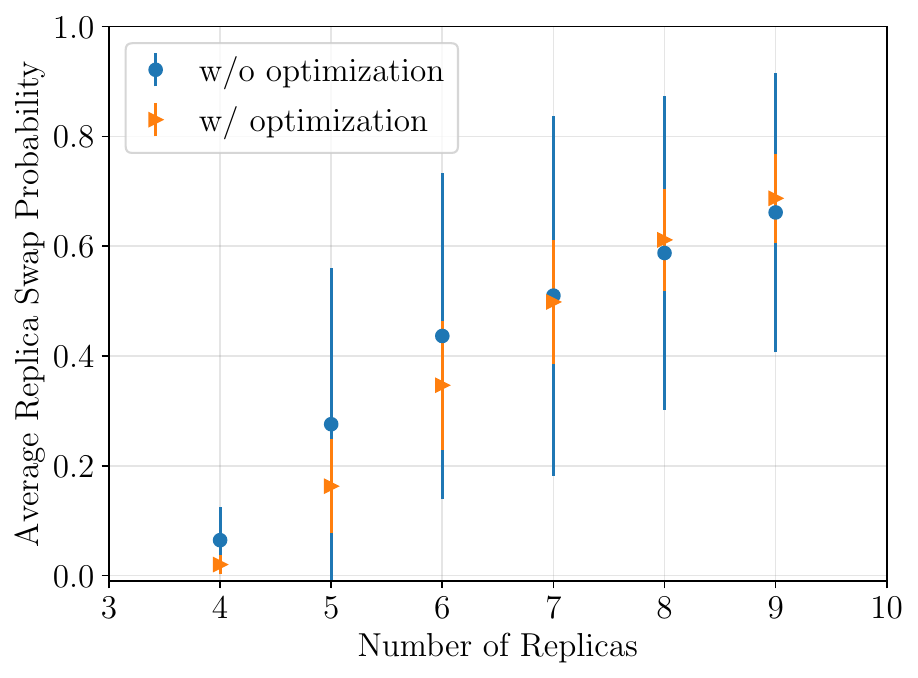}\label{fig:j1j2-replicasB}}
  \caption{(a) Relative frequency of finding an energy below the first excited state ($-34.70212$) of the open boundary condition 4$\times$5 $J_1$-$J_2$ problem when varying the number of replicas over $10^2$ independent runs, with the error bars corresponding to the $2 \sigma$ confidence interval calculated using twice the standard error of the mean. (b) The average swap probability between replicas at the end of each run. }
  \label{fig:j1j2-replicas}
\end{figure}

While optimizing temperatures of PT does lead to improved performance for the $J_1$-$J_2$ problem, the degree of improvement is highly dependent on the choice of hyperparameters. In this section, we further examine the 4$\times$5 $J_1$-$J_2$ model using an RBM architecture to determine the dependence of the final results on the hyperparameter choices.

We first explore how the PT algorithm performs as the number of samples used to estimate the next update increases in Fig.~\ref{fig:j1j2-samples}. As the success rate increases from 150 to 200 samples, the success rate steadily increases from $0.34 \pm 0.095$ to $0.64 \pm 0.096$ when the temperatures are allowed to optimize, and from $0.17 \pm 0.075$ to $0.30 \pm 0.091$ when the temperatures are held constant. 
The ratio of improvement remains about a factor of 2 in all of these cases.
However, the success rate peaks dramatically at 200 samples both with and without temperature optimization. The success rate for larger sample sizes tends to saturate with a rate of $0.30 \pm 0.091$ when optimizing the temperatures and $0.24 \pm 0.085$ without optimizing temperatures at 400 samples. We hypothesize that this peak at 200 samples is due to the finite samples providing beneficial noise that allows certain replicas to bump out of local minima. As the sample size increases ,this noise is reduced, and it becomes harder to escape the local minima.

Finally, we also investigate varying the number of replicas within the PT algorithm in Fig.~\ref{fig:j1j2-replicas}. We find a strong peak in the success rate when allowing the temperatures to optimize at 6 replicas. Beyond this point, the success rate generally seems to saturate near $0.60$ when the temperature is allowed to optimize. The success rate for cases where the temperature is not optimized also slowly increases to nearly match the adjusting temperature rate, demonstrating that as more temperatures are covered, the benefits of optimizing the temperature become less strong. 

However, we do see more variable success rate results around 4 and 5 replicas, with 4 replicas seeing a spike in success rate when temperatures are allowed to optimize. This may be related to an optimum average swap probability, as shown in Fig.~\ref{fig:j1j2-replicasB}. As the number of replicas increases, the average swap probability increases relatively linearly for both cases where the temperature is allowed to optimize and cases where it is not. The spread of the probabilities is much narrower when the temperatures are allowed to optimize, however, confirming that the optimization algorithm is leading to a degree of convergence in the swap probability. Probabilities around $0.20$ have previously been associated with optimum success probabilities in similar PT algorithms~\cite{Kone2005}. In our work, the optimum probability of swapping appears to be around $0.40$ for the 6 replica cases. This higher rate of swapping may be related to the limited number of samples and total updates performed in this work. 

Taken together, these results show that the overall success rate of our PT algorithm for ground state discovery is, to a certain extent, dependent on the hyperparameters used within the simulation. Nevertheless, in all cases studied, optimizing the temperature does improve the performance of the PT algorithm, or at least provides equal performance with minimal additional computational cost.
Furthermore, it can often significantly reduce the needed computational cost by providing high success rates at lower numbers of samples and replicas. 

\section{Energy Comparisons} \label{app:MeanEnergies}
For completeness, we show in Tables~\ref{tab:32_6x6}, \ref{tab:48_6x6}, and \ref{tab:64_6x6} the energies measured after $2\times10^4$ training steps of the feedforward networks from Sec.~\ref{sect:FFN} for PT with temperature optimization, PT without temperature optimization, and no PT. The energies of the successful runs, defined as the subset of runs that found an energy lower than the first excited state energy, are consistent within error bars in all three cases. There is, however, a large improvement over the no PT case for the PT with and without temperature optimization cases if we consider the mean over all runs, indicating that PT on average achieves lower energies.

These results imply that the main limitation to finding a lower energy is the expressivity of the network. What PT and temperature optimization offer is the ability to more often reach the limit of the architecture's expressivity. This is seen most clearly in Tables~\ref{tab:48_6x6} and \ref{tab:64_6x6} where the total and success means for PT with temperature optimization are identical. In this case, the training procedure found a minima lower than the first excited state by the $2\times 10^4$ step in every simulation we performed.
\begin{table}[!h]
\begin{tabular}{|l|l|l|}
\hline
& Total Mean               & Success Mean   \\ \hline
No-PT  & $-64.3 \pm 0.9$   & $-70.23 \pm 0.03$  \\ \hline
PT-NTO& $-69.9 \pm 0.1$    &  $-70.244 \pm 0.007$ \\ \hline
PT-TO& $-70.15 \pm 0.05$   &  $-70.253 \pm  0.005$ \\ \hline
\end{tabular}
\caption{Mean energies for the (32,16,16) network on the $6\times 6$ $J_2-J_2$ model with $J_2=0.62$ over 50 independent runs. We compare no parallel tempering (No-PT) to parallel tempering without temperature optimization (PT-NTO) to parallel tempering with temperature optimization (PT-TO).}\label{tab:32_6x6}
\end{table}

\begin{table}[!htb]
\begin{tabular}{|l|l|l|}
\hline
& Total Mean                    & Success Mean   \\ \hline
No-PT & $-66.4 \pm 0.6$          & $-70.29 \pm 0.02$ \\ \hline
PT-NTO& $-70.0 \pm 0.1$     &  $-70.314 \pm 0.009$ \\ \hline
PT-TO& $-70.309 \pm 0.008$     &   $-70.309 \pm 0.008$ \\ \hline
\end{tabular}
\caption{Mean energies for the (48,24,24) network on the $6\times 6$ $J_2-J_2$ model with $J_2=0.62$ over 50 independent runs. We compare no parallel tempering (No-PT) to parallel tempering without temperature optimization (PT-NTO) to parallel tempering with temperature optimization (PT-TO).}\label{tab:48_6x6}
\end{table}

\begin{table}[!htb]
\begin{tabular}{|l|l|l|}
\hline
& Total Mean                    & Success Mean   \\ \hline
No-PT & $-67.5 \pm 0.4$          & $-70.30 \pm 0.02$ \\ \hline
PT-NTO& $-70.17 \pm 0.07$     &  $-70.36 \pm 0.01$ \\ \hline
PT-TO& $-70.36 \pm 0.008$     &   $-70.36 \pm 0.008$ \\ \hline
\end{tabular}
\caption{Mean energies for the (64,32,32) network on the $6 \times 6$ $J_2-J_2$ model with $J_2=0.62$ over 50 independent runs. We compare no parallel tempering (No-PT) to parallel tempering without temperature optimization (PT-NTO) to parallel tempering with temperature optimization (PT-TO).}\label{tab:64_6x6}
\end{table}

\clearpage
\end{document}